\documentclass[12pt]{iopart}
\usepackage{graphicx}
\usepackage{color,amstext}
\begin{document}

\title[Angular characteristics of MM-fiber SPR sensor ]{Angular characteristics of a multimode fiber surface plasmon resonance sensor under wavelength interrogation}

\author{Zhixin Tan$^{1,2,3,4,*}$, Xin Hao$^4$, Xuejin Li$^{3,5,\ddag}$, Yuzhi Chen$^{3,5}$, Xueming Hong$^{3,4}$, Ping Fan$^{3,4}$}

\address{$^1$ Institute of High Energy Physics, Chinese Academy of Sciences
(CAS)，Beijing 100049, China}
\address{$^2$ Dongguan Neutron Science Center, Dongguan 523808, China}
\address{$^3$ Shenzhen Key Laboratory of Sensor Technology, Shenzhen, Guangdong,  China ~518060}
\address{$^4$ College of Physics Science and Technology, Shenzhen University, Guangdong, China ~518060}
\address{$^5$ College of Electronic Science and Technology, Shenzhen University,  Guangdong, China ~518060}

\ead{tanzhixin@ihep.ac.cn \hspace{2em} lixuejin@szu.edu.cn}

\begin{abstract}
  In this paper the angular characteristics of a multimode fiber SPR sensor are theoretically investigated. By separating the contributions of beams incident at different  angles, a compact model is presented to predict the shift of the resonance  wavelength with respect to the angle and the environmental refractive index. The result suggests that  the performance of conventional fiber SPR sensors  can be substantially improved by optimizing the incident angle. Furthermore, our investigation suggests some problems in previous reports.
\end{abstract}


\submitto{\JPD}

\maketitle{}

\section{Introduction}

Recent applications of fiber SPR (Surface Plasmon Resonance) sensors to biochemical analysis have attracted much interest~\cite{pollet2009,delport2012realtime,Knez2014bios,japan2010,jpd2012,lee2012OL,raey2015rev,tan2014plas,Tan2014}.   
By integrating SPR region on the cylindrical surface, a compact multimode fiber SPR sensor is produced. Fiber SPR sensor works based on attenuated total reflection (ATR) spectroscopy. For wavelength interrogation, a modification of surface mass load on the fiber sensor will change the resonance condition, which results in a spectral shift of the resonance dip. Compared with other variants, multimode fiber SPR sensors are simple, robust and easy to produce~\cite{Wang201388,Perrotton2013,tan2012spie}.  
To provide for a growing need in this developing field, we built a sensor system and created software for real-time measurement~\cite{tan2012spie}.  In such a conventional fiber SPR sensor, broadband light in various incident angles result in an asymmetric absorption curve, which requires high order polynomial fitting to resolve the resonance wavelength.This implies that it is possible to improve sensor performance by restricting the angular distribution of input light to a narrow range.

Incident angle is an important parameter in surface plasmon resonance. It determines the projection of the wave vector at the interface. A multimode fiber with a diameter of hundreds of microns is rigid enough to keep straight and to preserve the propagation angle of a light beam. This is the experimental basis of our investigation. Although there are already many theoretical and experimental studies of multimode fibers~\cite{Sharma2004423,sharma2007review,kanso2008}, an analysis to distinguish  the contributions of light beams with various incident angles is still missing. Meanwhile, the community lacks a clear picture for sensor prediction and optimization. This motivates us to re-visit this subject from a theoretical perspective. 
In this paper we focus on the angular characteristics of the multimode fiber SPR sensor under wavelength interrogation. The main framework is based on Kreschmann's theory and Fresnel reflections in multilayer films.

\section{The calculation model}
\begin{figure}[htb]
  \centering
  \includegraphics[width=0.85\textwidth]{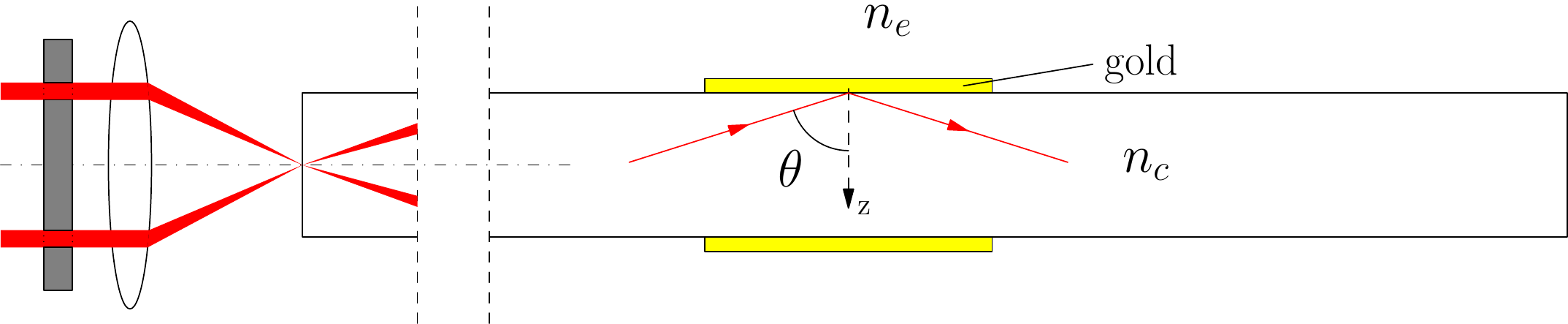}
  \caption{A multimode fiber SPR sensor with controlled light angle.}
  \label{fig:kreschmann}
\end{figure}

The schematic of a multimode fiber SPR sensor is shown in Fig.~\ref{fig:kreschmann}. The light is launched through a ring mask for angular control. Collimated beams are then focused at the center of the input face of the fiber with a microscope objective. Assuming the fiber sensor is fabricated from a plastic cladding  silica fiber with diameter $d=600~\mu m$, the removed cladding length is $L=4~mm$ and the thickness of the gold film is $t=45~nm$.  

Considering a circularly polarized light source, its transmission as a function of incidence angle and wavelength is given by~\cite{Sharma2004423,sharma2007review}:
\begin{eqnarray}
  \label{eq:trans}
  T(n_e, \theta, \lambda) &=& \frac{1}{2} R_{p}^N(n_e, \theta,  \lambda) + \frac{1}{2}R_{s}^N(n_e, \theta, \lambda) 
\end{eqnarray}
where $N =  L/d \tan \theta$ is the number of reflections. The transmitted light is divided into two components, with p- and s-polarization. The quantum factor of the CCD is not included since beams are considered in one angle with same efficiency.

The cascaded reflectance in three-layer films is given as follows~\cite{Sharma2004423}:
\begin{equation}
  \label{eq:reflection}
  R = \left|\frac{r_{cm} + r_{me} \exp{(2ik_{mz}t)}}{1+r_{cm}r_{me} \exp{(2ik_{mz}t})}\right|^2
\end{equation}
where the fiber-core, metal and environmental layers are denoted by c, m, and e,
respectively.  The fiber-core/metal and metal/environment interfaces are labeled as `cm' and `me'. The reflection coefficients with s- and p- polarized light on the interface are given for wave vectors along the z-axis and dielectric constants~\cite{Sharma2004423,sharma2007review}:
\begin{eqnarray}
  \label{eq:rpcm}
  r_{cm}^{p} &=& \frac{k_{cz}/\varepsilon_c - k_{mz}/\varepsilon_m}{k_{cz}/\varepsilon_c +
               k_{mz}/\varepsilon_m}    \\
\label{eq:rpme}
  r_{me}^{p} &=& \frac{k_{mz}/\varepsilon_m - k_{ez}/\varepsilon_e}{k_{mz}/\varepsilon_m +
               k_{ez}/\varepsilon_e}   \\
  r_{cm}^{s} &=& \frac{k_{cz} - k_{mz}}{k_{cz} + k_{mz} }  \\
  r_{me}^{s} &=& \frac{k_{mz} - k_{ez}}{k_{mz} + k_{ez} } 
\end{eqnarray}
The refractive indices of the fiber core are interpolated from ~\cite{silica1965}, while the dielectric constants of gold are according to Johnson and Christy~\cite{PhysRevB}. 
Wave vectors in different mediums can be written as:
\begin{eqnarray}
  \label{eq:zx}
  k_{iz}  &=& (\varepsilon_i \frac{\omega^2}{c^2} - k_x^2) ^{1/2} \\
  k_x &=& n_c  \frac{2 \pi}{\lambda_0}   sin \theta 
\end{eqnarray}
where $c$ is the speed of light and $\lambda_0$ is the wavelength of the incident beam in free space. The letter $i$ can be c, m, or e depending on the medium in question. $k_x$ is the component of the wave vector parallel to the interface, which is conserved in all media. In principle, the incident angle of the light beam  $\theta$ is bounded by the critical angle $\theta_{cr}=\sin^{-1}(n_{cl}/n_c)$ and $\pi/2$.
The fiber numerical aperture $ \mathrm{NA} = \sqrt{n_c^2 - n_{cl}^2}$ is also related to these indices. For a  multimode fiber with a numerical aperture NA=0.50, the critical angle is $69.8^\circ$ if $n_c = 1.446$~RIU is used. In our investigation, we discuss incident angles ranging from $65^\circ$ to $90^\circ$.

\begin{figure*}[htb]
  \centering
  \includegraphics[width=1.1\textwidth]{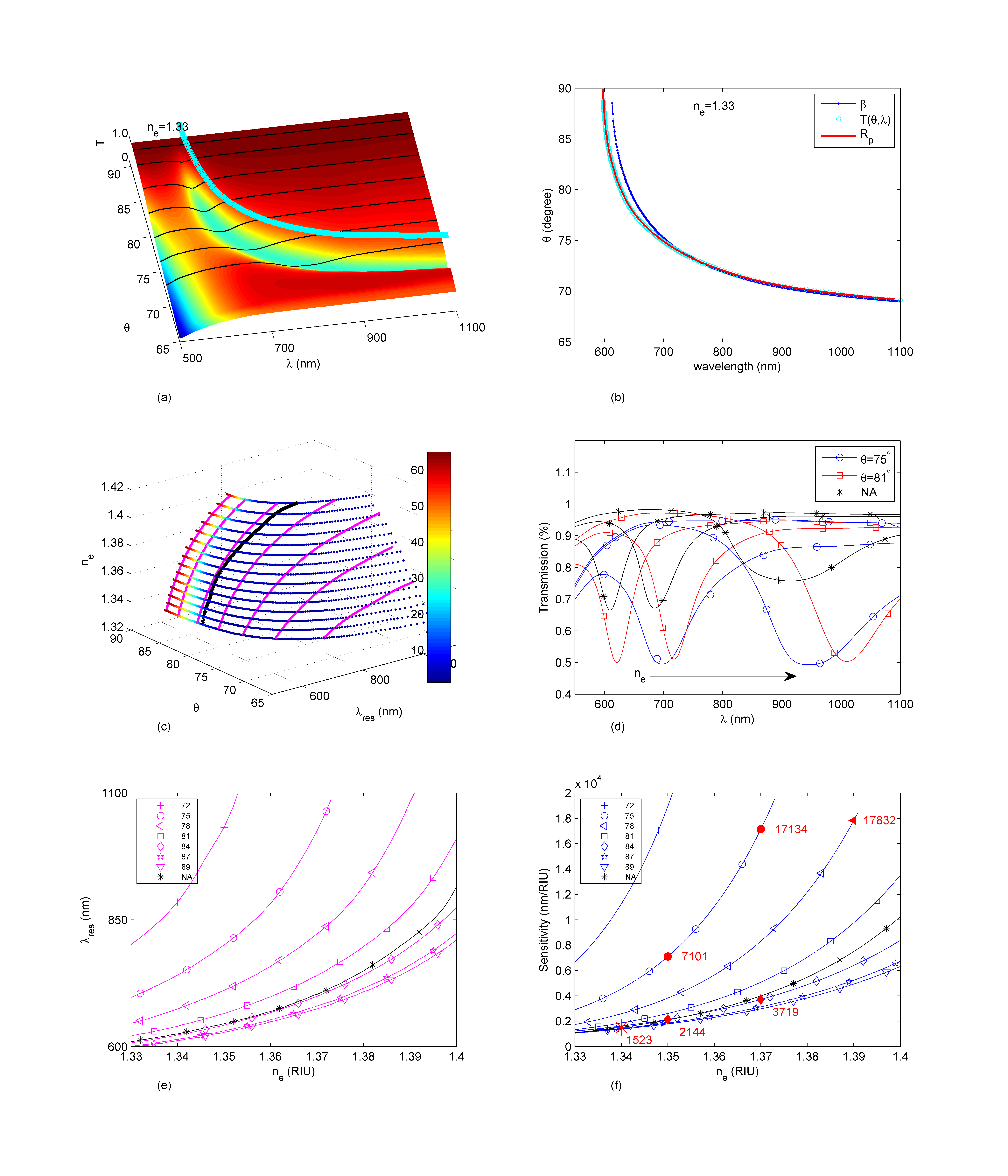}
  \caption{(a) Three-dimensional plot of the transmission ratio as a function of both wavelength and incident angle, for $n_e=1.33$~RIU. (b) The  ideal resonance requirement (blue), the general trend from the full model (cyan), and the result for a single reflectance $R_p$ (red). (c) The stacked resonance positions for $n_e$ ranging from 1.33 RIU to 1.40~RIU. The black dots are the projected working position of a conventional fiber sensor. (d) Sample spectra for beams with incident angles $\theta=81^\circ$ (red) and $\theta=75^\circ$ (blue), as well as the conventional fiber sensor (black).  (e) Resonance wavelength versus environmental index for a range of specific values for the incident angle. (f) Sensitivity versus environmental index for the same incident angles as (e).}
  \label{fig:3dlayer}
\end{figure*}

\section{Results and discussions}

Assuming the  fiber SPR sensor is immersed in the water ($n_e=1.33$~RIU), Fig.~\ref{fig:3dlayer} (a) presents a three-dimensional plot of the transmission ratio  $T=T(1.33, \theta,  \lambda)$   with respect to both the angle and the wavelength. For specific values of the incident angle ($\theta$ ranging from $72^\circ$  to $87^\circ$ with an interval of $3^\circ$, as well as $89^\circ$), black lines outline their absorption spectra. The absorption peak positions (corresponding to minimum transmission) are traced by a cyan line to show the overall tendency. Clearly, decreasing the incident angle increases  the resonance wavelength.  
The reason for this trend is not straightforward, since the surface plasmon vector $k_{sp}$ changes as the light wavelength increases. To explain this, we define an auxiliary parameter $\beta(\lambda_0) = \varepsilon_m(\lambda_0) \varepsilon_e/(\varepsilon_m(\lambda_0) + \varepsilon_e )$, which is a function of the wavelength $\lambda_0$.  To satisfy the resonance requirement, the component of the wave vector along the interface should be equal to that of the surface plasmon:
\begin{eqnarray}
  k_x &=& k_{sp} \label{eq:reso}   \\
  n_c \frac{2\pi}{\lambda_0} \sin \theta &=&  \frac{2 \pi}{\lambda_0} \sqrt{\beta(\lambda_0)}
\end{eqnarray}
 Thus the resonance position is defined as a relationship between the angle and the wavelength.
\begin{equation}
 \theta = \sin^{-1}({\sqrt{\beta(\lambda_0)}/n_c }) 
 \label{eq:sati}
\end{equation}
In Fig.~\ref{fig:3dlayer} (b), the blue line presents resonance points obtained from Eq.~\ref{eq:sati}, while the cyan line is the locus of peak absorption from Fig.~\ref{fig:3dlayer}(a). The two curves do not perfectly overlap, since they arise from different models; the blue line is derived assuming an ideal interface between two semi-infinite media, while the cyan line represents a working sensor with a thin film. The discrepancy between the two models can be accounted for by invoking an `effective' dielectric constant.
As the metal film thickness increases, the negative effective complex dielectric constant $\varepsilon_{\text{eff}}$ on the metal side of the interface decreases, which is equivalent to the increment of $n_e$ for they are equal in the expression of $\beta$. Thus the resonance wavelength increases slightly and the cyan line moves toward the blue line. In theory, the sensor interface approaches to the ideal model as the film thickness increases. 

To investigate the performance of the fiber sensor in different environmental conditions, resonance lines for $n_e$ from 1.33~RIU to 1.40~RIU are stacked in Fig.~\ref{fig:3dlayer} (c). The z-axis is the environmental refractive index. The lines are colored according to the transmission ratios at the resonance positions, as indicated by the color-bar. The heavy magenta lines are contours of constant incidence angle. These same magenta lines are plotted in Fig.~\ref{fig:3dlayer} (e) for clarity. In Fig.~\ref{fig:3dlayer}(f)  we present the corresponding sensitivities, which is defined as the ratio of the shift of the resonance wavelength to the change of environmental refractive index. For our case, we fit the discrete series first, and then calculate the corresponding value of the derivative of the fitted function. The meaning of the heavy black line is discussed below. Furthermore, Fig.~\ref{fig:3dlayer}(d) presents the transmission ratios for two specific values of the incident angle, $\theta=81$ and $\theta=75$, to give a direct impression of their shapes and FWHMs.

After processing, we achieve a systematic understanding for the performance of a multimode fiber SPR sensor. In particular, Fig.~\ref{fig:3dlayer} (c) presents an elegant perspective for sensor optimization, and from it we can draw three conclusions. First, the working angle is limited; it ranges from  $69^\circ$ to $90^\circ$ in our setup. 
Therefore, there is no justification for using a base fiber with a numerical aperture larger than 0.5.
Second, lower angles $\theta$ give higher sensitivity. The tradeoff is that its measurement range is narrower than that with large angles. 
To obtain high sensitivity, we should move the working position to a magenta line towards the right side of Fig~\ref{fig:3dlayer} (c). For example, when $n_e =1.35$~RIU,  the sensitivity is 2144~nm/RIU for a light beam with angle $\theta=84^\circ$ , while for a beam with $\theta=75^\circ$ (near the edge of a multimode fiber with NA=0.39), the sensitivity can be improved to 7101~nm/RIU, almost triple that for $\theta=84^\circ$.  These values are highlighted as red points in Fig.~\ref{fig:3dlayer} (f). 
For an even higher environmental index $n_e=1.37$~RIU, the corresponding sensitivities are 17134~nm/RIU and 3719~nm/RIU, differing by a factor of five. Third, when the selection of the incident angle is adapted to the environmental refractive index range to be  measured, the maximum sensitivities under different conditions are close to each other since their working positions are located on the right edge of the spherical face. For $n_e = 1.39$ RIU, the corresponding sensitivity is 17832~nm/RIU if a light beam with $\theta=78^\circ$ is used. This result is close to the red point of 17134~nm/RIU under $n_e=1.37$~RIU in Fig.~\ref{fig:3dlayer} (f).

To compare our analysis with previous investigations, we model a multimode fiber SPR sensor with incoming light comprising a distribution of incident angles, corresponding to a numerical aperture NA=0.22. This model is henceforth referred to as a `conventional fiber sensor'. Other parameters are the same as for the previously discussed single-angle model. Transmission ratios and resonance wavelengths for different $n_e$ are calculated, taking into account the angular dependence of the power distribution and the CCD efficiency~\cite{Sharma2004423,sharma2007review,kanso2008}. To allow direct comparison, we introduce a concept of ``effective angle'', whose value is  interpolated from the resonance wavelength of the corresponding resonance line. Thus we  project the result of the conventional fiber sensor onto the spherical face in Fig.~\ref{fig:3dlayer} (c), as illustrated by the heavy black line.  As shown, the black line lies almost on top of the magenta line corresponding to $\theta=84^\circ$, so the effective angle of the conventional fiber sensor is about $\theta=84^\circ$. 
The black line lies entirely in the blue (high absorption) region, so the red (low absorption) zone near the edge of the plot, which corresponds to light beams in angles greater than $87^\circ$, is inactive in practice.  It suggests that light beams being almost parallel to the fiber axis, corresponding to the red edge of the surface in Fig.~\ref{fig:3dlayer} (c), contribute little to the sensor performance because of their weak absorption.
As $n_e$ increases, the resonance wavelength increases and the wave vector $k_x = k \sin\theta$ decreases; this leads to preferential weighting of smaller angles, which should account for the deviation of the black line towards a smaller effective angle at large refractive indices in Fig.~\ref{fig:3dlayer}(c).

The transmission curves for a conventional fiber sensor for environmental indices $n_e=1.33$~RIU, $n_e=1.365$~RIU, and  $n_e=1.40$~RIU are plotted in Fig.~\ref{fig:3dlayer} (d) with legend `NA'. The resonance absorptions are less than 0.4, which suggests a moderate resonance state caused by the short SPR region.
In general, transmission for the conventional fiber sensor is a superposition of resonance curves of various angles.   For the condition $n_e =1.40$~RIU, the FWHM of the spectrum is close to 200~nm and the bottom of the curve is nearly flat, which makes it hard to identify the minimum position. 
Compared to the red lines ($\theta = 81^\circ$), the FOM of the sensor is much poor.
In Fig.~\ref{fig:3dlayer} (e) and (f) we show resonance wavelengths and sensitivities of the conventional fiber sensor  (black line and star markers). These results are somewhat higher than for light with a single incidence angle of $\theta=84^\circ$.  
The sensitivity  is 1523~nm/RIU at $n_e = 1.34$~RIU, as labeled in (f). This value is close to what we found in our experimental study~\cite{tan2012spie}, if we neglect some differences in sensor parameters. 

Except the low efficiency and moderated performance in the conventional configuration, many earlier reports did not take a serious consideration for the angular characteristics of the fiber SPR sensor~\cite{tan2012spie,pollet2009,delport2012realtime,japan2010}. For example, the Y-combiner, which is widely found in the literature and involves thermal tapering and then fusing together of fibers, will change the angular distribution of the propagating light and the re-distribution in angle is out of control. Therefore it is impossible to trace the paths of rays. But SPR effect strongly depends on the incident angle.  Thus it leads to a mess in theory. More attention should be paid to this issue. 
Based on our investigation, we recommend that researchers working on multimode fiber SPR sensors stick to the original configuration without Y-combiner, although it is a bit inconvenience.
Another important problem is the mismatch in aperture size between the fiber itself and the spectrometer, which had been neglected in both the experiment and its theoretical simulation. For compactness and convenience, most fiber SPR configurations use a fiber spectrometer to acquire spectra. However, the fiber spectrometer has its own numerical aperture, typically NA = 0.22.  Obviously, under this configuration the large-angle contributions are wasted, even though the base fiber has a large NA.  A possible solution is to use another microscopic objective to recover a quasi-collimated light beam before it enters the spectrometer; otherwise one could use a customized fiber spectrometer with a larger NA. 

About twenty years ago several fiber SPR sensors with oblique light input were built for angular interrogation~\cite{gupta1994,Mono1996,Trouillet1996}. Obviously, their configurations were much different from our proposal as we use white light and wavelength interrogation while laser and angle shift are adopted in those reports. The kernel of our proposal is to purify the input and evaluate the contribution separately. This can not be achieved in these mentioned papers where non-meridian light beam dominate the effect. In addition, if the ring mask is thick enough, the skew rays should be reduced greatly in our configuration. 

The number of reflections $N(\theta)$ is determined by the fiber diameter and the sensing length. Since power operation doesn't change the position of the minima, these two parameters have little effect on our model if the absorption threshold hasn't been reached. This is also demonstrated in Fig.~\ref{fig:3dlayer} (b). The red line, which represents the result of a single reflection of $R_p$ in Eq.~(\ref{eq:trans}), overlaps almost exactly with the cyan line of minimum transmission; the small difference is a numerical artifact arising from the minimum search algorithm. 
In fact, if the averaging reflection times is fixed as N=1, the SPR model reduces to a prism SPR model. In this case Fig.~\ref{fig:3dlayer} (c) maintains its shape, and corresponding conclusions still hold. Thus our model may also serve as a reference for prism SPR studies.

When the sensing length increases, $N(\theta)$ increases, the resonance is intensified and the red zone in Fig.~\ref{fig:3dlayer}(c) shrinks to the edge. Consequently,  the importance of light beam with large angle increases; working positions of the conventional fiber SPR sensor move toward the edge and the sensitivity decreases. This should account partly for differences between our numerical results and previous reports~\cite{tan2012spie,kanso2008}.

As illustrated in Fig.~\ref{fig:3dlayer}(b), the film thickness make a slight difference between a fiber SPR sensor of finite thickness and the ideal model. If the film thickness changes, the cambered surface in Fig.~\ref{fig:3dlayer} (c) shifts slightly. Furthermore, the sensitivity is determined as a slope of the magenta curves. Thus the variation in performance is delicate and it is difficult to identify a general tendency. In contrary, the film thickness plays an important role in the signal amplitude, which should be considered before other factors.

Fiber SPR tips or tapered fiber SPR can also be described qualitatively in our model. As proposed in works ~\cite{Verma20081486,Yuan2013}, post-processing of the fiber will directly decrease the incident angle of the main portion of the light with respect to the normal of the cone surface. Thus the effective working angle decreases and the sensitivity is improved. Meanwhile, our model helps to identify the maximum angle for post-processing: according to Fig.~\ref{fig:3dlayer} (b), the tapering angle should be less than $20^\circ$ under environmental index $n_e = 1.33$~RIU.

\section{Conclusions}

In conclusion, we have numerically calculated and analyzed the angular characteristics of a multimode fiber SPR sensor. A general model is presented for sensor prediction and optimization. It suggests that small-angle components of the light beam have better sensitivity. We have proposed an angular fiber SPR sensor based on a multimode fiber.  For a specific measuring index, the performance of a multimode fiber SPR sensor can be substantially improved by deliberate control of the beam angle.

The author would like to thank Dr. Scott Edwards, Shenzhen University, for his kind help in language editing. The work was supported  by the National Science Foundation of China under Grant (No.61275125 and No.61308046), Basic Research Program of Shenzhen and High-level Talents Project of Guangdong Province.

\vspace{2em} 

\bibliographystyle{unsrt}

\end{document}